\documentclass[12pt,reqno]{article}
\usepackage{graphicx}
\usepackage{cite}
\usepackage{amssymb,amsmath,amscd,amsthm}
\usepackage{times}
\usepackage[bookmarks=false]{hyperref}
\hypersetup{%
    colorlinks=true,        
    linkcolor=blue,          
    citecolor=blue,         
    urlcolor=blue           
    }

\usepackage[left=2.54cm, right=2.54cm, top=2.54cm, bottom=2.54cm]{geometry}

\allowdisplaybreaks
\date{}

\title{An Oscillating Holographic Dark Energy in $f(R)$ Gravity}
\author{A. Y. Shaikh$^1$, D. P. Tadas$^{2}$\footnote{Corresponding author E-mail: dtadas144@rediffmail.com}, S. D. Katore$^3$ \\ 			
	$^1$Department of Mathematics, Indira Kala Mahavidyalaya Ralegaon, India.\\
	$^2$Toshniwal Arts, Commerce and Science College, Sengaon-431542, India.\\
	$^3$Department of Mathematics, Sant Gadge Baba Amravati University\\ Amravati-444602, India.}
\date{}
\begin{document}
\maketitle

\begin{abstract}
In this article, we investigated a Locally Rotationally Symmetric (LRS) Bianchi-II cosmological model with matter and Holographic Dark Energy (HDE) in the context of {$f(R)$} theory of gravity. In order to find exact solutions to the field equations, we assumed that the Shear scalar $(\sigma )$ is proportional to Expansion scalar $(\theta)$. For HDE, it is observed that the Equation of state (EoS) parameter $ \omega_{\Lambda } $ has an oscillating nature and lies in $[-0.778,\, 1.016].$ Also, we have studied the validity of energy conditions and shows that Null Energy Condition (NEC) is violated near the bouncing points. Moreover, we analysed the physical and geometrical aspects of the investigated model.

\end{abstract}


\section{Introduction}
The recent astronomical studies, including High-Redshift Supernovae, SNe Ia, Galaxy clustering, Cosmic microwave background and Large-scale structure, have confirmed that the universe is expanding at an accelerated rate \cite{r01,r02,r03,r04,r05}. To explore the accelerating and expanding nature of the universe, one approach is to modify general theory of relativity (GTR). Several modified theories of GTR such as  $f(R)$, $f(T)$, $f(R,T)$, $f(G)$ \cite{r06,r07,r08,r09,r10,r11} and so on, have been proposed to rectify the problems in cosmology. The simplest modifications to GTR are the $f(R)$ theories, which introduces the arbitrary function of Ricci scalar in Einstein-Hilbert action. Firstly, Buchdahl \cite{r12} have proposed the $f(R)$ theory of gravity to generalize Einstein's GTR. The modified theories have considered as a very interesting theories to describe the dark energy. The new class of modified $f(R)$ gravity have proposed by Nojiri and Odintsov \cite{r13} which unify $R^{m} $ early-time inflation with late-time $\Lambda CDM$ epoch. Wu and Zhu \cite{r14} has reconstructed the $f(R)$ theory which effectively describes the HDE and reconstruct the function $f(R)$ with the parameter $c<1, c=1$ and $c>1$ respectively. Many features of $f(R)$ theories and various applications to cosmology and gravity, such as dark energy, inflation, local gravity constraints, cosmological perturbations, and spherically symmetric solutions in weak and strong gravitational backgrounds, have been studied by Felice and Tsujikawa \cite{r15}. Katore \cite{r16} have investigated Bianchi-II, VIII and IX cosmological models with string in the framework of $f(R)$ theory of gravitation. Shamir \cite{r17} has investigated an anisotropic LRS Bianchi-I metric in  $f(R)$ theory of gravity. Also, Santhi et al. \cite{r18} have investigated LRS Bianchi-I with bulk viscous string cosmological model in $f(R)$ gravity and shows that the realistic energy conditions, $\rho \ge 0$ and $\rho _{p} \ge 0$ are satisfied. Recently, An anisotropic cosmological model with magnetized strange quark matter have investigated by Ozdemir and Aktas \cite{r19} in $f(R)$ theory of gravity.  Also, Katore and Gore \cite{r20} have studied $\Lambda$CDM Bianchi-I cosmological models with HDE in $f(R)$ gravity theory.

Another alternative for solving the dark energy problem is HDE, which is based on the Holographic concept initially presented by G't Hooft \cite{r21} in the context of black hole physics. Without decaying into a black hole, this principle should be limited by an infrared cutoff scale $ L $ and ultraviolet cutoff scale $\Lambda $. Fischer and Susskind \cite{r22} have proposed the cosmological version of holographic principle and have been discussed some of its consequences. The model of HDE is the idea of the short distance cut-off relates to the infrared cut-off has proposed by Li \cite{r23}. Setare and Vagenas \cite{r24} investigated the cosmic dynamics of the interacting HDE model. Adhav et al. \cite{r25} have investigated the homogeneous and anisotropic Bianchi-I cosmological model with interacting dark matter and HDE. Granda and Rojas \cite{r26} have studied an analysis of the autonomous system, corresponding critical points and their stability using HDE in the context of $f(R)$ gravity. Recently, Hatkar et al. \cite{r27} have investigated the Bianchi-III cosmological model with viscous HDE in Brans-Dicke theory of gravity. Shaikh and his collaborators \cite{r28} have studied HDE in $f(G)$ gravity.

The homogeneous and anisotropic Bianchi-II cosmological model plays a vital role in describing the universe at the early stages of its evolution. The homogeneous but anisotropic LRS Bianchi-II perfect fluid cosmological model with constant deceleration parameter have studied by Singh and Kumar \cite{r29}. Total energy of the universe has obtained by Aydogdu \cite{r30} based on LRS {\tiny }Bianchi-II models by using the Tolman, Papapetrou, and Weinberg energy-momentum complexes. Amirhashchi \cite{r31} has explored a novel anisotropic LRS Bianchi-II cosmological model using stiff fluid with variable $\Lambda$. Tiwari et al. \cite{r32} have studied the LRS Bianchi-II cosmological model in the presence of perfect fluid with varying $\Lambda$. Adhav \cite{r33} has used exponential and power-law expansions to find solutions to field equations for the LRS Bianchi-II cosmological model with anisotropic dark energy. Reddy and Santhi \cite{r34} have investigated the LRS Bianchi-II perfect fluid metric in the $f(R,T)$ theory of gravity. Sarkar \cite{r35} has investigated the correspondence between LRS Bianchi-II interacting HDE and tachyon scalar field dark energy model with variable deceleration parameter. Maurya et al. \cite{r36} have studied anisotropic LRS Bianchi-II DE models with a new approach of time-dependent deceleration parameter. R. L. Naidu et al. \cite{r37} have studied anisotropic Bianchi-V dark energy cosmological model with massive scalar field. Mete et al. \cite{r38} have examined the LRS Bianchi-II space time with interacting dark matter and HDE for $n=\frac{1}{2} $and $n=\frac{3}{2} $. Recently, Pradhan et al. \cite{r39,r40} have studied the LRS Bianchi-II cosmological models in GTR and the modified gravity presented by Saez–Ballester.

With the observation of above discussion, we have studied LRS Bianchi-II HDE cosmological models in the context of $ f(R) $ gravity theory and explored the possibilty of bouncing solutions. In section 2, we derived the $ f(R)$ gravity field equations for the LRS Bianchi-II cosmological model in the presence of HDE. The scale factor solutions in respect to the field equations determined in section 3. The section 4, devoted to analysed the physical and geometrical parameters. The energy conditions and its graphical interpretaion are discussed in section 5. The last section is concluded with discussion and conclusion.

\section{Metric and Field Equations}
The simplest and most popular modification of GTR is the $f(R)$ theory of gravity. For the $f(R)$ gravity, the modified action is given as follows:
\begin{equation}\label{Eq:1}
	S=\int \sqrt{-g} \left(f(R)+L_{m} \right)\,  d^{4} x, 
\end{equation} 
where $f(R)$ is an arbitrary function of the Ricci scalar $R$ and $L_{m} $ is the matter Lagrangian. It is worth mentioning that the standard of Einstein-Hilbert action can be recovered when $f(R)=R.$

Now, the $f(R)$ gravity field equations are obtained by varying the action with respect to metric $g_{\mu \nu } $ are as follows:
\begin{equation}\label{Eq:2}
	F(R)R_{\mu \nu } -\frac{1}{2} f(R)g_{\mu \nu } -\nabla _{\mu } \nabla _{\nu } F(R)+g_{\mu \nu } \nabla ^{\mu } \nabla _{\mu } F(R)=-(T^{m} {}_{\mu \nu } +T^{\Lambda } {}_{\mu \nu } ), 
\end{equation}
where $F(R)\equiv \frac{df(R)}{dR}$, $\nabla _{\mu }$ denotes covariant differentiation, $T^{m} {}_{\mu \nu } $ and $T^{\Lambda } {}_{\mu \nu } $ are the energy momentum tensor corresponding to the matter and HDE respectively.

We have considered the LRS Bianchi-II space-time in an orthogonal frame given by \cite{r31} as
\begin{equation}\label{Eq:3}
	ds^{2} =-dt^{2} +A^{2} (dx^{2} +dz^{2} )+B^{2} (dy+xdz)^{2} , 
\end{equation} 
where  $ A $ and $ B $ are functions of time $ t $ only.

The scalar curvature $ R $ for the metric \eqref{Eq:3} is given by 
\begin{equation}\label{Eq:4}
	R=-2\, \left[\frac{2A_{44} }{A} +\frac{B_{44} }{B} +\frac{2A_{4} B_{4} }{AB} +\frac{A_{4} {}^{2} }{A^{2} } -\frac{B^{2} }{2A^{4} } \right],
\end{equation} 
where the subscript '4' denotes differentiation with respect to time $ t $.

The energy momentum tensors for matter and HDE are defined as
\begin{equation}\label{Eq:5}
	T^{m} {}_{\mu \nu } = \rho _{m} u_{\mu } u_{\nu } ,  T^{\Lambda } {}_{\mu \nu } = (\rho _{\Lambda } +p_{\Lambda } )u_{\mu } u_{\nu } + g_{\mu \nu }  p_{\Lambda } , 
\end{equation} 
where $\rho _{m} ,$ $\rho _{\Lambda } $ are the energy densities of matter and HDE respectively and $p_{\Lambda } $ is the pressure corresponding to HDE.

\noindent The field equations can be expressed as follows using equations \eqref{Eq:3}, \eqref{Eq:4} and \eqref{Eq:5}.
\begin{equation}\label{Eq:6}
	F\left[\frac{2A_{44} }{A} +\frac{B_{44} }{B} \right]+\frac{1}{2} f(R)+\frac{2A_{4} F_{4} }{A} +\frac{B_{4} F_{4} }{B} =-(\rho _{m} +\rho _{\Lambda } ),
\end{equation}
\begin{equation}\label{Eq:7}
	F\left[-\frac{A_{44} }{A} -\frac{A_{4} {}^{2} }{A^{2} } -\frac{A_{4} B_{4} }{AB} +\frac{B^{2} }{2A^{4} } \right]-\frac{1}{2} f(R)+\frac{3A_{4} F_{4} }{A} +\frac{B_{4} F_{4} }{B} +F_{44} =-p_{\Lambda } ,
\end{equation}
\begin{equation}\label{Eq:8}
	F\left[-\frac{B_{44} }{B} -\frac{2A_{4} B_{4} }{AB} -\frac{B^{2} }{2A^{4} } \right]-\frac{1}{2} f(R)+\frac{2A_{4} F_{4} }{A} +\frac{2B_{4} F_{4} }{B} +F_{44} =-p_{\Lambda } ,
\end{equation}
\begin{equation}\label{Eq:9}
	F\left[-\frac{A_{44} }{A} -\frac{A_{4} {}^{2} }{A^{2} } -\frac{A_{4} B_{4} }{AB} +\frac{B^{2} }{2A^{4} } \right]+\frac{[AA_{4} +BB_{4} x^{2} ]\, F_{4} }{A^{2} +B^{2} x^{2} } -\frac{1}{2} f(R)+\frac{2A_{4} F_{4} }{A} +\frac{B_{4} F_{4} }{B} +F_{44} =-p_{\Lambda } ,
\end{equation}
where the derivative with respect to cosmic time $ t $ is denoted by the subscript '4'.

The equation of HDE density is given by
\begin{equation}\label{Eq:10}
	\rho _{\Lambda } =3(\alpha H^{2} +\beta H_{4} ), 
\end{equation} 
where $\alpha , \beta $ are constants and $H$ is the Hubble parameter. This condition, proposed by Granda and Oliveros \cite{r41}, which must fulfill the limitations imposed by current observations.

The continuity equations for matter and HDE can be obtained as
\begin{equation}\label{Eq:11}
	(\rho _{m} )_{4} +(\rho _{\Lambda } )_{4} +\left(\frac{2A_{4} }{A} +\frac{B_{4} }{B} \right)(\rho _{m} +\rho _{\Lambda } +p_{\Lambda } )=0, 
\end{equation}

The equation of continuity for the matter is
\begin{equation}\label{Eq:12}
	(\rho _{m} )_{4} +\left(\frac{2A_{4} }{A} +\frac{B_{4} }{B} \right)\rho _{m} =0, 
\end{equation}

The equation of continuity for the HDE is
\begin{equation}\label{Eq:13}
	(\rho _{\Lambda } )_{4} +\left(\frac{2A_{4} }{A} +\frac{B_{4} }{B} \right)(\rho _{\Lambda } +p_{\Lambda } )=0. 
\end{equation}

The equation of state for HDE is defined as 
\begin{equation}\label{Eq:14}
	\omega _{\Lambda } =\frac{p_{\Lambda } }{\rho _{\Lambda } } . 
\end{equation}

Using equations \eqref{Eq:10} and \eqref{Eq:13} in \eqref{Eq:14}, the EoS parameter becomes
\begin{equation}\label{Eq:15}
	\omega _{\Lambda } =-1-\frac{2\alpha HH_{4} +\beta H_{44} }{3H(\alpha H^{2} +\beta H_{4} )} . 
\end{equation} 
\section{Solutions of Field Equations}
The field equations \eqref{Eq:6}-\eqref{Eq:9} are a system of four equations in six unknowns $A$,  $ B $, $ f $, $ \rho _{m} $, $ \rho _{\Lambda } $ and $ p_{\Lambda } $. To find the solutions of the field equations, we consider the relations as:
\begin{enumerate}
	\item  We suppose that the Shear scalar $\sigma $ is proportional to Expansion scalar $\theta $, resulting in a relation between scale factors $ A $ and $ B $ as
	\begin{equation}\label{Eq:16}
		B=A^{n},
	\end{equation}
	where $n$ is an arbitrary real number.
	\item  Secondly, for the simplicity, we assume the technique of Boutrous \cite{r42} and Ram \cite{r43} and the adhoc relation as 
	\begin{equation}\label{Eq:17}
		\frac{A_{44} }{A} +\frac{A_{4} {}^{2} }{A^{2} } +\frac{A_{4} B_{4} }{AB} -\frac{B^{2} }{2A^{4} } =\frac{B_{44} }{B} +\frac{A_{4} B_{4} }{AB} +\frac{B^{2} }{2A^{4} } . 
	\end{equation} 
\end{enumerate}

Using equations \eqref{Eq:16} and \eqref{Eq:17}, we have obtained as
\begin{equation}\label{Eq:18}
	A_{44} +(n+1)\frac{A_{4} {}^{2} }{A} =\frac{A^{2n-3} }{1-n} ;\quad n\ne 1. 
\end{equation}

We denote, $A_{4} =S(A)$ which implies that $A_{44} =SS',$ where $S'=\frac{dS}{dA}.$ Therefore, after solving and integrating above equation leads to
\begin{equation}\label{Eq:19}
	S^{2} =\frac{2}{4n(1-n)} A^{2n-2} +c_{1} {}^{2} \frac{2}{(1-n)} A^{-2(n+1)} ;\quad n\ne 1. 
\end{equation}

We assume that $n=2$ to find the deterministic solution in cosmic time.  Therefore, above equation reduce to
\begin{equation}\label{Eq:20}
	\frac{2A^{3} dA}{\sqrt{4c_{1} {}^{2} -A^{8} } } =dt. 
\end{equation}

On integrating above equation, we get
\begin{equation}\label{Eq:21}
	A=\left(2c_{1} \sin (2t+c_{2} )\right)^{\frac{1}{4} } . 
\end{equation}

Therefore, using equations \eqref{Eq:16} and \eqref{Eq:21}, we get
\begin{equation}\label{Eq:22}
	B=\left(2c_{1} \sin (2t+c_{2} )\right)^{\frac{1}{2} } . 
\end{equation} 
where $c_{1} $ and $c_{2} $ are integrating constants.

It is clear from equations \eqref{Eq:21} and \eqref{Eq:22} that, both the metric potentials $A$ and $ B $ have oscillating nature and has physical meaning with Piao and Zhang \cite{r45}.  Hence using equations \eqref{Eq:21} and \eqref{Eq:22}, the LRS Bianchi-II space time for HDE can be written as
\begin{equation}\label{Eq:23}
	ds^{2} =-dt^{2} +\sqrt{2c_{1} \sin (2t+c_{2} )} (dx^{2} +dz^{2} )+\left(2c_{1} \sin (2t+c_{2} )\right)\, (dy+xdz)^{2} . 
\end{equation} 

\section{Physical Parameters}

We define the Scale factor $ a(t) $ and mean Hubble's parameter $H$ for the LRS Bianchi-II metric \eqref{Eq:3} may be defined as
\begin{equation}\label{Eq:24}	
	a(t)=(A^{2} B)^{\frac{1}{3}}
\end{equation}
\begin{equation}\label{Eq:25}
	H=\frac{1}{3} \frac{V_{4} }{V} =\frac{1}{3} \left(\frac{2A_{4} }{A} +\frac{B_{4} }{B} \right), 
\end{equation}
where $V=A^{2} B$ is the spatial volume of the universe.\

Let us first define the dynamical scalars, such as the expansion scalar $\theta$, shear scalar $\sigma^2$, and mean anisotropic parameter $\Delta _m$:
\begin{equation}\label{Eq:26}
	\theta =3H, 
\end{equation} 
\begin{equation}\label{Eq:27}
	\sigma ^{2} =\frac{1}{3} \left(\frac{A_{4} }{A} -\frac{B_{4} }{B} \right)^{2} , 
\end{equation} 
\begin{equation}\label{Eq:28}
	\Delta _{m} =\frac{1}{3} \sum _{i=1}^{3}\left(\frac{H_{i} -H}{H} \right)^{2}  , 
\end{equation} 
where $H_{i} \, \left(i=1,\, 2,\, 3\right)$ represents the directional Hubble parameter in the directions of $x,y,z$ respectively. The mean anisotropic parameter of the expansion $\Delta _{m} $, plays an important role in determining whether the model is isotropic or anisotropic. It is the measure of deviation from isotropic expansion; when $\Delta _{m} =0,$\ the universe expands isotropically.

Also, the deceleration parameter $q$ which is defined as
\begin{equation}\label{Eq:29}
	q=\frac{d}{dt} \left(\frac{1}{H} \right)-1. 
\end{equation}

The cosmological behavior of the universe is predicted by the sign of $q.$ For an accelerating universe, the deceleration parameter  $q$ is negative, while for a decelerating universe, it is positive.

Using euations \eqref{Eq:21} and \eqref{Eq:22}, the value of scale factor $ a(t) $ obtained as
\begin{equation}\label{Eq:30}
	a(t)=\left(2c_{1} \sin (2t+c_{2} )\right)^{\frac{1}{3} },
\end{equation}

The mean Hubble's parameter $H$ as well as the deceleration parameter $q$ are obtained as
\begin{equation}\label{Eq:31}
	H=\frac{2}{3} \cot (2t+c_{2} ) 
\end{equation} 
\begin{equation}\label{Eq:32}
	q=2+3\tan ^{2} (2t+c_{2} ), 
\end{equation} 
where $c_{2} $ is integrating constant.

\begin{figure}[htb]
	\centering
	\includegraphics[width=0.55\linewidth]{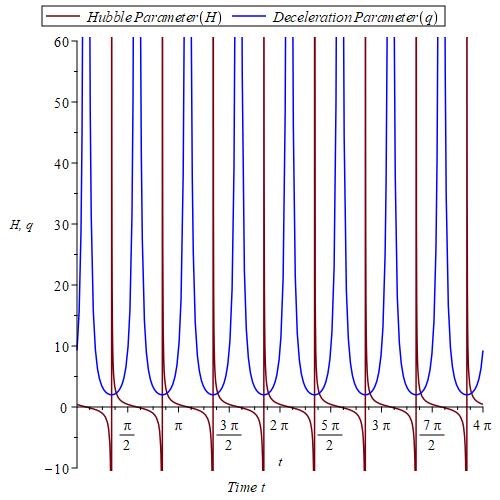}
	\caption{The plots of Hubble parameter $(H)$, Deceleration parameter $(q)$ \textit{vs.} time $t$ for $c_{2} =1.$}
	\label{f01}
\end{figure}

It is observed that, for this model the scale factor vanishes throughout each cycle and Hubble parameter diverges $ i.e. $ becomes singular at bouncing point. The bounce occurs when for $t=\left(\frac{n\pi }{2} -\frac{1}{2} \right)$, where $n$ is natural number which correspond to Big Crunch or Big bang singularity.  Also, the deceleration parameter $q>0$ for any $t\ge 0.$ Hence, the positive sign of $q$ indicates that investigated model undergoes in the phase of deceleration and its behavior is depicted in figure 1.

Also, the physical quantities Shear scalar $\sigma ^{2}, $ Expansion scalar $\theta $ and mean anisotropic parameter $\Delta _{m} $ are obtained as

\begin{equation}\label{Eq:33}
	\sigma ^{2} =\frac{3}{4} \cot ^{2} (2t+c_{2} ), 
\end{equation}
\begin{equation}\label{Eq:34}
	\theta =2\cot (2t+c_{2} ), 
\end{equation} 
\begin{equation}\label{Eq:35}
	\Delta _{m} =\frac{1}{8} . 
\end{equation}

It is clear from the above equations that, the mean anisotropic parameter $\Delta _{m} $ is independent of time $t$, $i.e.$ it remains constant throughout the universe's evolution. As a result, the current model appears to be anisotropic.
\begin{figure}[htb]
	\centering
	\includegraphics[width=0.55\linewidth]{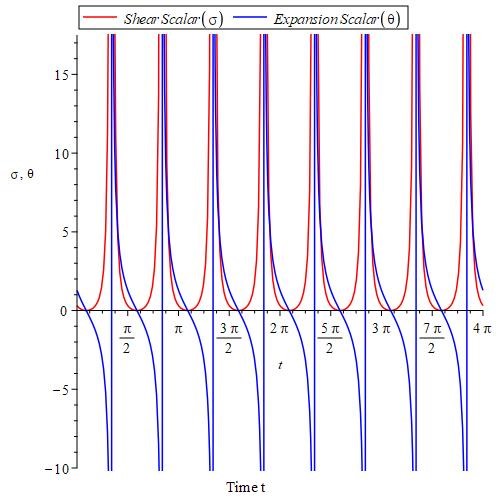}
	\caption{The plots of Shear scalar $(\sigma ^{2} )$ and Expansion scalar $(\theta )$ \textit{vs.} time $t$ for $c_{2} =1.$}
	\label{f02}
\end{figure}

Therefore, from equations \eqref{Eq:33} and \eqref{Eq:34}, we get
\begin{equation}\label{Eq:36}
	\frac{\sigma ^{2} }{\theta ^{2} } =\frac{3}{16} ={\rm constant.} 
\end{equation}

The ratio $\frac{\sigma }{\theta } ={\rm constant}$ implies that the model under investigation does not approach isotropy. For the given metric \eqref{Eq:3}, the energy densities for matter and HDE are obtained as
\begin{equation}\label{Eq:37}
	\rho _{m} =c_{3} \cdot \text{cosec}(2t+c_{2} ), 
\end{equation}
\begin{equation}\label{Eq:38}
	\rho _{\Lambda } =\frac{1}{4} \left[33+5\cot ^{2} (2t+c_{2} )\right]D-\frac{E}{2} -c_{3} \cdot \text{cosec}(2t+c_{2} ), 
\end{equation}
where $ D, $ $ E, $ $ c_{2} $ and $c_{3}$ are integrating constants.\
\begin{figure}[htb]
	\centering
	\includegraphics[width=0.55\linewidth]{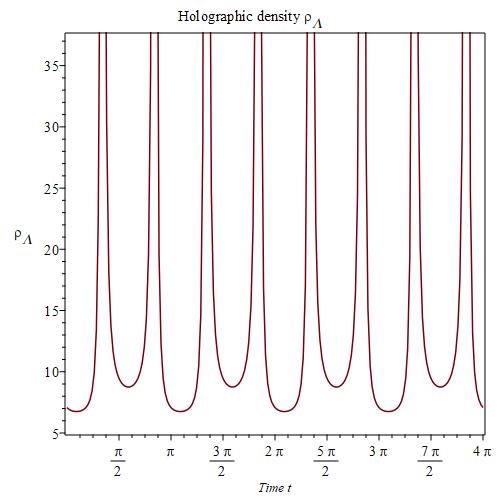}
	\caption{The graphical behaviours of Holographic density $(\rho _{\Lambda } )$ \textit{vs.} time $t$ for $D=E=c_{2} =c_{3} =1.$}
	\label{f03}
\end{figure}

The holographic density $(\rho _{\Lambda } )$ has oscillates in the positive quadrant as shown in figure 3. The density $\rho _{\Lambda } \ge 0$ is observed, indicating that the investigated model \eqref{Eq:23} satisfies the weak energy conditions. Our findings are similar to those of Beesham \cite{r46} and Rami \cite{r47}.

Also, the pressure for HDE is obtained as
\begin{equation}\label{Eq:39}
	p_{\Lambda } =\left[\frac{5}{4} \cot ^{2} (2t+c_{2} )-\frac{23}{4} \right]D+\frac{E}{2}.       
\end{equation} 
where $D$ and $E$ are constants of integration.

\begin{figure}[h]
	\centering
	\includegraphics[width=0.55\linewidth]{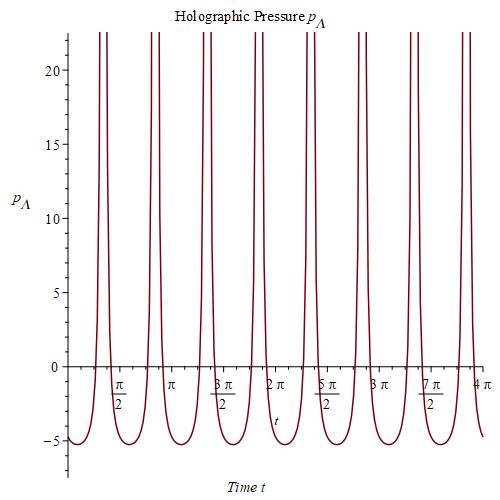}
	\caption{The graphical behaviours of Holographic Pressure $(p_{\Lambda } )$ \textit{vs.} time $t$ for $D=E=c_{2} =1.$}
	\label{f04}
\end{figure}

From figure 3 and 4, it is observed that, density $(\rho_{\Lambda } )$ and pressure $(p_{\Lambda } )$ of HDE diverges for $t=\left(\frac{n\pi }{2} -\frac{1}{2} \right)$, where $n$ is natural number  $ i.e. $ becomes singular at bouncing point. This singuarity occurs at bounce when the scale factor tends to zero  and density as well as pressure becomes infinite called Type III singularity. The asymptotic behaviour of type III singilarity has observed by Setare et al. \cite{r55} for additive $ g(T) $ solution.

From equations \eqref{Eq:14}, \eqref{Eq:38} and \eqref{Eq:39}, the HDE equation of state (EoS) parameter $\omega _{\Lambda } $ becomes

\begin{equation}\label{Eq:40}
	\omega _{\Lambda } =\frac{\left[\frac{5}{4} \cot ^{2} (2t+c_{2} )-\frac{23}{4} \right]D+\frac{E}{2} }{\left[\frac{33}{4} +\frac{5}{4} \cot ^{2} (2t+c_{2} )\right]D-\frac{E}{2} -c_{3} \cdot \text{cosec}(2t+c_{2} )} .
\end{equation}
\begin{figure}[h]
	\centering
	\includegraphics[width=0.55\linewidth]{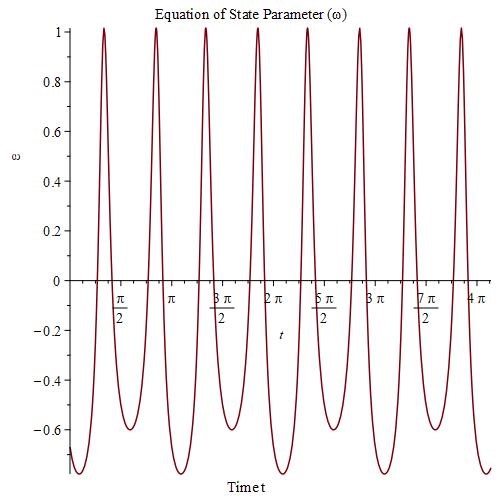}
	\caption{The physical behaviours of Equation of state (EoS) parameter $(\omega )$ \textit{vs.} time $t$ for $D=E=c_{2} =c_{3} =1.$}
	\label{f05}
\end{figure}

Figure 5 depicts the evolution of EoS parameter $\omega _{\Lambda } $ for HDE, which shows that it starts decreasing until it becomes negative, then increases until it becomes positive after a finite time and reverses back. As a result, the EoS parameter oscillates during the cycle resembles with \cite{r56} and varying between -0.778 and 1.016. $i.e.$ for any $t\ge 0\Rightarrow $ $\omega _{\Lambda } \in [-0.778,\, 1.016].$ Durrer and Laukenmann \cite{r44} have investigated the oscillating universe as a possible alternative to inflation..

Using equations \eqref{Eq:4}, \eqref{Eq:21} and \eqref{Eq:22}, then Ricci scalar becomes
\begin{equation}\label{Eq:41}
	R=\, \left[\frac{5}{2} \cot ^{2} (2t+c_{2} )-\frac{17}{2} \right]. 
\end{equation}

Solving equations \eqref{Eq:7}, \eqref{Eq:8}, \eqref{Eq:16} and \eqref{Eq:41}, we get
\begin{equation}\label{Eq:42}
	f(R)=\left[\frac{5}{2} \cot ^{2} (2t+c_{2} )-\frac{17}{2} \right]D+E, 
\end{equation} 
where $D$  and $E$ are integrating constants.

Also, the matter density parameter $\Omega _{m} $ and HDE density parameter $\Omega _{\Lambda } $ are defined and obtained as 
\begin{equation} \label{Eq:43}
	\Omega _{m}=\frac{\rho _{m}}{3H^{2}} =\frac{3c_{3} }{4} \tan (2t+c_{2} ) \sec (2t+c_{2} ) 
\end{equation}
and 
\begin{equation}\label{Eq:44}
	\Omega _{\Lambda } =\frac{\rho _{\Lambda}}{3H^{2}}=\left(\frac{99D}{16}-\frac{3E}{8}\right)\tan ^{2} (2t+c_{2} )+\frac{15D}{16}-\frac{3c_{3} }{4} \tan (2t+c_{2} )\sec (2t+c_{2} )
\end{equation}

Therefore, using equations \eqref{Eq:43} and \eqref{Eq:44}, the total energy density parameter $\Omega =\Omega _{m} +\Omega _{\Lambda } $ is obtained as
\begin{equation} \label{Eq:45}
	\Omega =\left(\frac{99D}{16}-\frac{3E}{8}\right)\tan ^{2} (2t+c_{2} )+\frac{15D}{16} 
\end{equation}
\begin{figure}[h]
	\centering
	\includegraphics[width=0.55\linewidth]{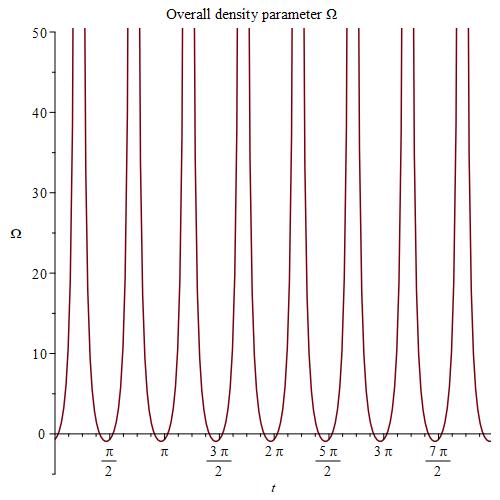}
	\caption{The plot of Overall density parameter  $\Omega $ \textit{vs.} time $t$.}
	\label{fig:06}
\end{figure}
From figure 6, we have observed that the overall density parameter  $\Omega $ oscillating during each cycle. During each cycle, it starts decresing from positive value to become negative and then goes incresing and become infinite. As a results, the values of overall density parameter  $\Omega <1$, $\Omega > 1$ and $\Omega = 1$ represents the open, closed and flat universe respectively.

\section{Energy Conditions}
The energy conditions are a set of linear equations derived from Raychaudhuri's equation that demonstrate that energy density can never be negative and gravity always attracts. The linear combination of pressure and density cannot be negative according to the energy conditions. Therefore, the energy conditions are expressed as follows:

\begin{enumerate}
	\item [(i)] 	NEC $\Leftrightarrow $ $\rho _{\Lambda } +p_{\Lambda } \ge 0$
	\item [(ii)]	SEC $\Leftrightarrow $ $\rho _{\Lambda } +3p_{\Lambda } \ge 0$
	\item [(iii)]	DEC $\Leftrightarrow $ $\rho _{\Lambda } >\left|p_{\Lambda } \right|\ge 0$ and
	\item [(iv)]	WEC $\Leftrightarrow $ $\rho _{\Lambda } \ge 0$.
\end{enumerate}

Using the equations  \eqref{Eq:37} and \eqref{Eq:38}, the expression for NEC and SEC  are obtain as
\begin{equation} \label{Eq:48} 
	\rho _{\Lambda } +p_{\Lambda } =\frac{5D}{2} \text{cosec}^{2} (2t+c_{2} )-c_{3} \text{cosec}(2t+c_{2} ) 
\end{equation} 
\begin{equation} \label{Eq:49}
	\rho _{\Lambda } +3p_{\Lambda } =5D\cot ^{2} (2t+c_{2} )+9D+E-c_{3} \text{cosec}(2t+c_{2} ) 
\end{equation} 
\begin{figure}[htb]
	\centering
	\includegraphics[width=0.49\linewidth]{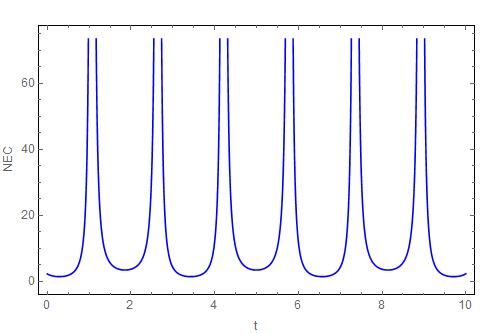}
	\includegraphics[width=0.49\linewidth]{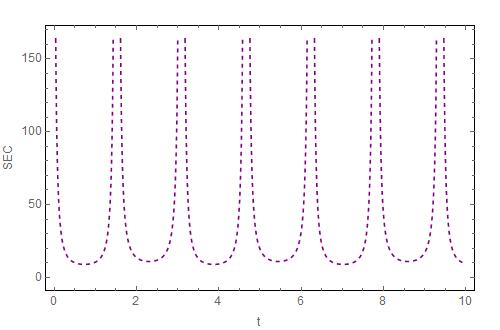}
	\caption{The plot of NEC and SEC \textit{vs.} time.}
	\label{f07}
\end{figure}
\begin{figure}[htb]
	\centering
	\includegraphics[width=0.49\linewidth]{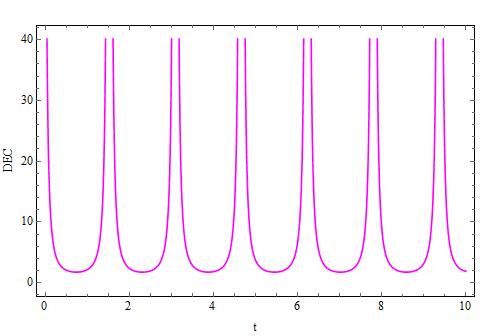}
	\includegraphics[width=0.49\linewidth]{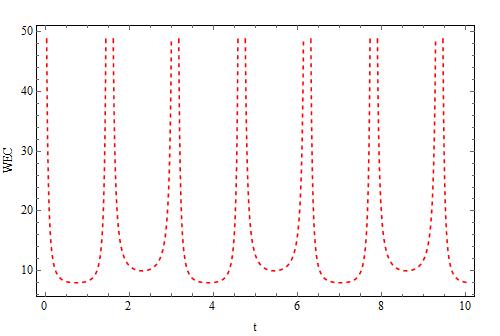}
	\caption[]{The plot of DEC and WEC \textit{vs.} time.}
	\label{f08}
\end{figure}

The graphical representations of NEC, SEC, DEC and WEC depicted in figure (7) and (8) shows that, they obeys the state of energy conditions. It's worth noting that our findings are more relevant than those of Ozdemir and Aktas \cite{r19} and Sahoo et al. \cite{r56}, because our model satisfies all of the energy conditions.

\section{Conclusions}
We have investigated the HDE cosmological model based on LRS Bianchi-II space time in the context of $f(R)$ gravity. The LRS Bianchi-II cosmological model in $f(R)$ theory of gravity has the solutions from equation \eqref{Eq:16}. It is observed that, our investigated model admits oscillating nature $i.e.$ it has cyclic nature. This investigated model represents the behaviour of a cyclic universe \cite{r49,r51} undergo as a sequence of continuous contraction and expansions phase \cite{r50,r52,r53,r54}. The mean anisotropic parameter $\Delta _{m} \ne 0$ which indicates that the investigated model is anisotropic throughout the evolution of the universe. Also, we have observed that the cosmological quantities like Shear scalar $\sigma $, Expansion scalar $\theta $ and Hubble parameter $H$ are oscillating infinitely and tends to infinity for $t=\left(\frac{n\pi }{2} -\frac{1}{2} \right)$ and all are vanishes at $t=\left(\frac{(2n-1)\pi }{4} -\frac{1}{2} \right)$, where $n$ is natural number. Also, the deceleration parameter $q>0$ implies that the decelerated expansion of the universe. It's important to note that our investigation resembles with \cite{r16,r46,r56}. Moreover, all of the energy conditions, including the NEC, SEC, DEC, and WEC, are also satisfied.

\end{document}